\documentstyle[aps,multicol]{revtex}
\input{epsf}

\begin{document}
\draft

\title{Iterated fully coordinated percolation on a square lattice}
\author{E. Cuansing and H. Nakanishi}
\address{Department of Physics, Purdue University, West Lafayette,
         Indiana 47907}
\date{\today}
\maketitle

\begin{abstract}

  We study, on a square lattice, an extension to fully coordinated
  percolation which we call iterated fully coordinated percolation.  In 
  fully coordinated percolation, sites become occupied if all four of its 
  nearest neighbors are also occupied.  Repeating this site selection process 
  again yields the iterated fully coordinated percolation model.  Our 
  results show a substantial enhancement in the size of highly connected 
  regions after each iteration (from ordinary to fully coordinated and then 
  to iterated fully coordinated percolation).  However, using Monte Carlo 
  methods to determine the static critical exponents and normal mode analyses 
  to determine the dynamic critical exponents, our results indicate that the 
  ordinary, fully coordinated, and iterated fully coordinated percolation 
  models belong to the same universality class.  This is in contrast to our
  previous results \cite{fully} wherein a different universality class
  {\em dynamically} (but not {\em statically}) for fully coordinated
  percolation was indicated, though the differences in the dynamic exponents
  were small.  A possible cause might be the different methods of generating
  clusters, i.e., in this work clusters were generated statically in square
  lattices while in our previous work clusters were grown dynamically, for
  studying the dynamic critical exponents.

  \medskip
  \noindent{PACS number(s): 
                 64.60.Ak,
                 64.60.Fr,
                 05.40.Fb,
                 05.70.Fh}

\end{abstract}

\begin{multicols}{2}
\narrowtext

\section{INTRODUCTION}
\label{intro}

\begin{figure}
  \epsfxsize=1.00\hsize \mbox{\hspace*{-0.065\hsize} \epsffile{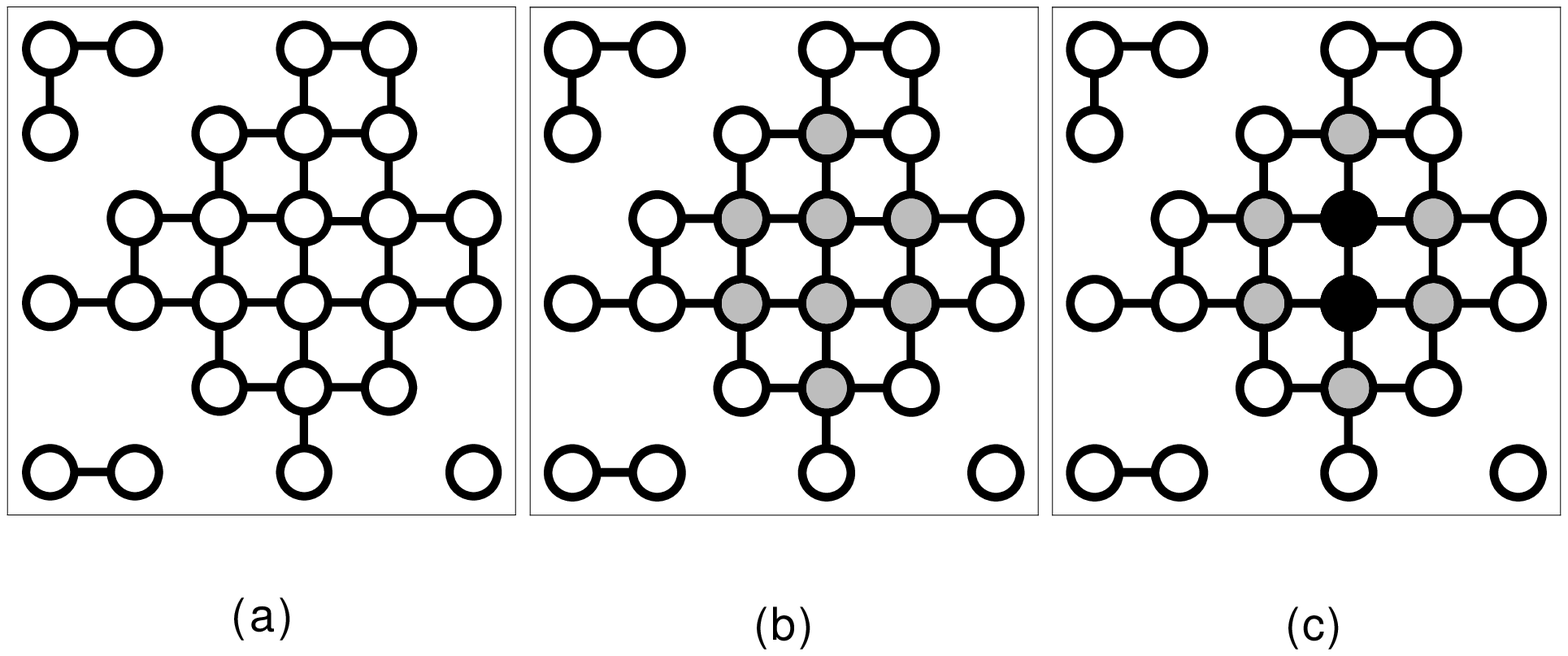}}
  \vspace{0.3cm}
  \caption{An illustration of the process to simulate IFC percolation. 
           (a) First we simulate ordinary percolation on a square lattice.
           Occupied sites are shown as circles while unoccupied sites are
           not shown.  (b) From the resulting sites from ordinary percolation,
           choose the fully coordinated sites, i.e., occupied sites whose four 
           neighbors are also occupied.  These sites are shown as gray-filled 
           circles.  (c) Repeat the culling of non-fully coordinated sites one 
           more time.  The remaining sites are then considered as sites in IFC
           percolation.  These sites are shown as solid circles in the figure.}
  \label{sim_stages}
  \end{figure}

  {\it Iterated fully coordinated percolation} (referred to as IFC percolation
  from this point on) is an extension to the ordinary percolation and the
  {\it fully coordinated percolation} (referred to as FC percolation hereon)
  models.  These are well-known problems of geometrical phase transition
  where global connectivity is singular as a function of local connectivity
  (see, for a review, Stauffer and Aharony \cite{stauffer94}). In ordinary
  site percolation on a square lattice, sites are randomly occupied with an
  independent occupation probability $p$. Two nearest neighbor sites are
  considered connected if they are both occupied. A collection of connected
  sites, i.e., every site in the group is connected to every other site either
  directly or through the paths of nearest neighbors, is termed a cluster.
  When $p$ is increased until it reaches a critical value $p_c$, an infinitely
  spanning cluster will form for the first time. The square lattice shown in 
  Fig.~\ref{sim_stages}(a) is an example of ordinary site percolation on a 
  square lattice.  Occupied sites are shown as circles while unoccupied sites 
  are not shown.

  To perform IFC percolation, first perform ordinary percolation.  After this
  is done, determine which of the occupied sites have all four of their
  nearest neighbors also occupied.  Occupied sites having this property are
  called {\it fully coordinated sites} and will remain occupied while the other
  sites will be removed.  This process of winnowing occupied sites was 
  described in \cite{fully} and we called the resulting model the
  {\it fully coordinated percolation} model (FC percolation). The square
  lattice shown in Fig.~\ref{sim_stages}(b) is an example of fully coordinated 
  percolation constructed from the ordinary percolation sites shown in 
  Fig.~\ref{sim_stages}(a).  The gray-filled circles are fully coordinated
  sites.  The final step in the process is to again choose fully coordinated 
  sites but this time from the resulting system constructed from FC 
  percolation.  As in the FC percolation case, we also remove the sites that 
  are not fully coordinated.  The solid circles in the square lattice shown in 
  Fig.~\ref{sim_stages}(c) are examples of IFC percolation sites constructed 
  from Figs.~\ref{sim_stages}(a) and \ref{sim_stages}(b).

  Notice that this problem is different from bootstrap percolation (see, for
  a review, Adler \cite{adler91}) where sites of less connectivity are
  iteratively removed until no more sites have insufficient connectivities.  
  This is in contrast to our problem where we only perform the culling of 
  non-fully coordinated sites twice.

  In our earlier work \cite{fully}, we found that the sizes of the 
  fully-connected subregions (with no internal holes) in an incipient 
  infinite cluster in FC percolation are much larger than those found in 
  ordinary percolation.  In this study, we find this same increase in the 
  sizes of the fully-connected subregions to continue when the full 
  coordination requirement is iterated one more time.  The regions of high 
  local connectivity are generally interesting because they would seriously 
  affect the vibrational modes in an equivalent disordered network of 
  {\em springs}.  In particular, it was found previously that they enter into 
  the scaling of the normal mode spectrum \cite{mukherjee96}.

  In the same previous work \cite{fully}, we also concluded that the dynamic
  critical exponents of FC percolation were close to but different from
  those of ordinary percolation, drawing this conclusion from numerical
  results on the FC percolation clusters which were grown from a seed site
  until prescribed sizes were reached.  In this work, on the other hand, we
  obtain numerical results on the corresponding dynamical critical exponents
  of FC and IFC percolation from the clusters generated statically on square
  grids of prescribed sizes.  Surprisingly, our new results indicate that
  the numerical values of these exponents are very similar for ordinary, FC,
  and IFC percolation, and in fact most likely to be the same.  Since the
  actual numerical estimates for FC cases are compatible within the error
  estimates of both works, this might simply mean that we drew an overzealous
  conclusion in our previous work.  However, we cannot rule out the possibility
  that the ensembles of clusters used in \cite{fully} and the current work
  may contain subtle differences which in fact alter dynamic universality
  classes.

  The rest of this paper is divided into four sections.  In 
  Sec.~\ref{exponents} we present numerical results and calculations of the 
  static and dynamic critical exponents of IFC percolation.  The geometry of 
  clusters is studied in Sec.~\ref{geometry} with a particular focus on the 
  sizes and distributions of the highly connected subregions.  The difference 
  between the methods of generating the cluster ensemble in \cite{fully} and 
  the current work is discussed in Sec.~\ref{comparison}.  The final section 
  gives a summary of our results.

\section{STATIC AND DYNAMIC CRITICAL BEHAVIOR}
\label{exponents}

  We have determined the static and dynamic critical exponents of IFC 
  percolation by simulations on a square grid.  The culling of non-fully
  coordinated sites occurs in two stages.  After sites in the grid are 
  randomly occupied with an independent probability $p$, all the fully
  coordinated sites are marked. After this marking process is done, all 
  the occupied but unmarked sites are removed.  From the resulting system
  of remaining sites, we again determine which sites are fully coordinated
  and remove those which are not.  Figs.~\ref{sim_stages} depict this process.
  To determine the connectivities among the sites, we have used the standard
  Hoshen-Kopelman algorithm \cite{hoshen76}.

  \begin{figure}
  \epsfxsize=0.90\hsize \mbox{\epsffile{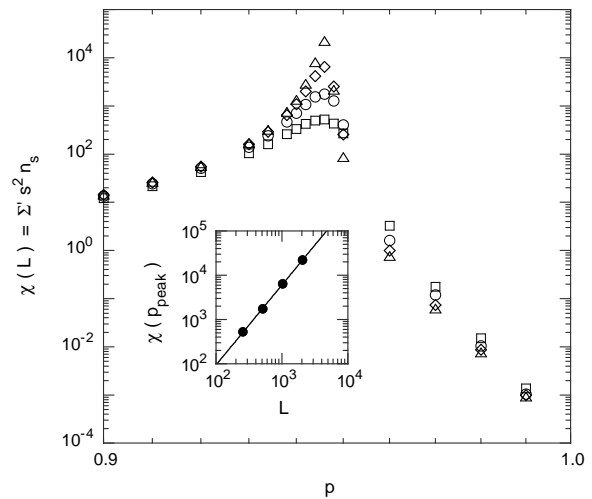}}
  \vspace{0.3cm}
  \caption{Susceptibilities $\chi$ against probability $p$ in IFC percolation
           on square lattices of sizes $L^{2}$ are shown.  The squares,
           circles, diamonds, and triangles are for $L = 256$, $512$, $1024$, 
           and $2048$, respectively.  Each data point in the plot is an average
           over $1000$ realizations.  Shown as the inset is the finite-size
           scaling of the susceptibility peaks by plotting the four peaks
           against $L$ on a log-log scale.  Note that $L$ is a dimensionless
           quantity corresponding to the length, in multiples of the lattice
           constant, of the side of the square lattice.}
  \label{susceptibilities}
  \end{figure}

  \vspace{0.1in}

  The critical occupation probability, $p_c$, is determined from the plot
  of susceptibility, $\chi$, against $p$.  This plot is shown in
  Fig.~\ref{susceptibilities}.  Notice that because of finite size effects,
  instead of a diverging $\chi$ at $p_c$, we see clearly defined peaks at $p$
  values near the true $p_c$.  The lattice sizes, $L^2$, we used in our
  simulations were $L = 256$, $512$, $1024,$ and $2048$, and each data point
  in Fig.~\ref{susceptibilities} is an average over $1000$ realizations.
  According to the plot, the susceptibility peaks are pegged at
  $p_{peak} = 0.946$ for the larger lattices.  Since the larger the
  lattices the more our simulations mimic the asymptotic limit, we can
  conservatively assign the value for the critical occupation probability
  to be at $p_c = 0.946 \pm 0.001$.

  \vspace{0.2in}

  \begin{table}
  \caption{The ratio between static critical exponents $\gamma$ and $\nu$, 
           and the fractal dimension $d_f$ found for ordinary, FC and IFC 
           percolation models are shown.  The values at the bottom row of the 
           table are previously determined values for ordinary percolation.  
           The $\gamma/\nu$ ratio for ordinary percolation is not determined
           in this work.}
  \label{tab:static}
  \vspace{0.1in}
  \begin{tabular}{ccc}
    percolation type & $\gamma/\nu$ & $d_f$ \\
    \tableline
    ordinary &        ---        & $1.888 \pm 0.004$ \\
    FC       & $1.791 \pm 0.006$ & $1.885 \pm 0.014$ \\
    IFC      & $1.805 \pm 0.018$ & $1.857 \pm 0.007$ \\
    \tableline
    ordinary &                   &                   \\
    (prior results) & $\frac{43}{24}\ (\approx 1.7917)$ 
    \tablenote{see, e.g., Stauffer and Aharony \cite{stauffer94}.} & 
    $\frac{91}{48}\ (\approx 1.8958)^{\ \rm a}$ \\
  \end{tabular}
  \vspace{0.1in}
  \end{table}

  The ratio between critical exponents $\gamma$ and $\nu$ can be determined
  from the scaling of susceptibility peak values against the corresponding
  lattice sizes, i.e., $\chi(p_{peak}, L) \sim L^{\gamma / \nu}$.  Using this
  relationship we find $\gamma / \nu$ for IFC percolation to be 
  $1.805 \pm 0.018$.  This value agrees nicely with the ordinary percolation 
  value which is known exactly to be $43/24 \approx 1.7917$.  The 
  ratio $\gamma / \nu$ can also be used to indirectly determine the fractal 
  dimension, $d_f$, from the known scaling relationship, 
  $d_f = 1 + \gamma / 2\nu$.  The result for IFC percolation is 
  $d_f = 1.903 \pm 0.009$ while the known exact value for ordinary percolation 
  is $91/48 \approx 1.8958$.  The critical exponent $\gamma$ can also 
  be determined independently from the scaling of susceptibility against the 
  occupation probability $p$, which goes as 
  $\chi(p,L=\infty) \sim |p - p_c|^{-\gamma}$, where the $L = \infty$ 
  indicates the scaling to be ideally satisfied in the asymptotic limit of 
  $L \rightarrow \infty$.  As in the case of FC percolation, the 
  susceptibility peaks are very close to $p = 1.0$.  This provides data to 
  the right of the peaks to reside in a small interval.  We therefore only use 
  data to the left of the peaks to determine the scaling implied.  The value 
  we found for $\gamma$ in IFC percolation using $p_c = 0.946$ is 
  $\gamma = 2.398 \pm 0.039$.  This value also agrees very well with the 
  value in ordinary percolation, known to be exactly $43/18 \approx 2.3889$. 

  The fractal dimension, $d_f$, of IFC percolation can also be determined
  directly.  At $p_c$, the size of the largest cluster, $S$, scales with $L$
  as $S \sim L^{d_f}$, where $d_f = 1.857 \pm 0.007$.  This value is close to 
  the known exact value for ordinary percolation as well as to the value
  obtainable from our measured $\gamma / \nu$.

  Listed in Table~\ref{tab:static} are the values for the ratio $\gamma/\nu$
  and for $d_f$ we found from our simulations of ordinary, FC, and IFC 
  percolation.  As a reference, the previously determined values of the 
  corresponding critical exponents for ordinary percolation are listed at the 
  bottom row of the table.

  The values for the spectral dimension, $d_s$, and the walk dimension, $d_w$,
  can be determined from the dynamic critical behavior of IFC percolation.
  We first construct the transition probability matrix {\bf W}.  The elements,
  $W_{ij}$, of this matrix are the hopping probabilities per step of a 
  random walker to hop from site $j$ to site $i$.  The diagonal elements 
  $W_{ii}$ are the probabilities for the walker to stay at site $i$ at a time 
  step.  In the {\it myopic ant} rule, all of the $W_{ii}$ are zero.  In
  contrast, in the {\it blind ant} rule, the $W_{ii}$ are nonzero, their values
  dependent on how many occupied neighboring sites the walker can hop into 
  (see, for a review,\cite{nakanishi94}).  We are using the blind ant rule for 
  this work.

  Once the {\bf W} matrix is determined, the value of $d_w$ can be estimated
  directly from its largest non-trivial eigenvalue, $\lambda_1$.  This 
  eigenvalue satisfies the finite size scaling law \cite{nakanishi93}
  \begin{equation}
    |ln \lambda_1|~\approx~1 - \lambda_1~\sim~S^{-2/d_s}.
  \label{eigenval}
  \end{equation}
  Making use of the Alexander-Orbach scaling law \cite{alexander82},
  $d_s = 2 d_f/d_w$, and $S \sim L^{d_f}$, we obtain
  \begin{equation}
    1 - \lambda_1~\sim~L^{-d_w}.
  \label{walk_dim}
  \end{equation}
  Eq.~(\ref{walk_dim}) gives a rather direct way of estimating $d_w$ from
  finite size scaling, in a manner that is considerably simpler compared to
  methods wherein a detailed knowledge of the eigenvectors is required
  (see, e.g., \cite{mukherjee94}).

  \begin{figure}
  \epsfxsize=0.90\hsize \mbox{\epsffile{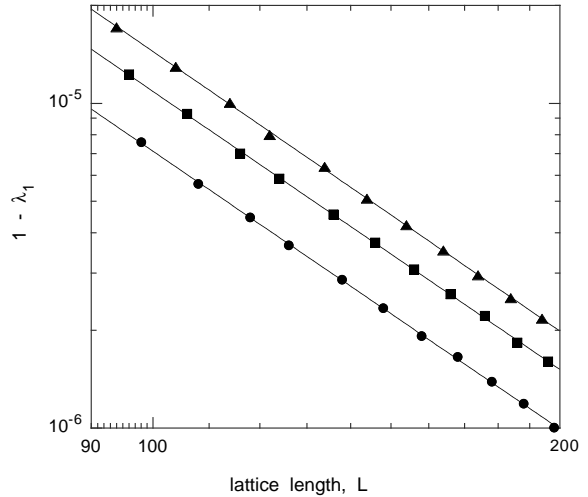}}
  \vspace{0.3cm}
  \caption{Shown is the scaling of the largest nontrivial eigenvalue
           $\lambda_{1}$ by plotting $1 - \lambda_{1}$ against the lattice
           length $L$ on a log-log scale.  The circles, squares and 
           triangles are for ordinary, FC, and IFC percolation, respectively.
           Each data point in the plot is an average over $1000$ random
           realizations.}
  \label{lambda}
  \end{figure}

  \vspace{0.1in}

  Shown in Fig.~\ref{lambda} is our data plotted in a way that is consistent 
  with Eq.~(\ref{walk_dim}).  For each percolation type, we used $13$ different 
  edge lengths of the square lattices that range from $L=94$ to $L=198$.  
  Furthermore, for each data point, about $1000$ random realizations of the 
  {\bf W} matrix had been generated by Monte Carlo simulations.  To 
  numerically determine $\lambda_1$ from each {\bf W} matrix, we used the 
  {\it Arnoldi-Saad} method \cite{saad91}.

  \vspace{0.2in}

  \begin{table}
  \caption{Shown are the walk and spectral dimensions for ordinary, FC, and
           IFC percolation.  The walk dimensions were determined by taking
           the largest non-trivial eigenvalue of {\bf W} while the spectral
           dimensions are derived from $d_s = 2 d_f/d_w$, with $d_f$ values
           taken from Table~I.  The last row in the table 
           contains previously known values for ordinary percolation.}
  \label{tab:dynamic}
  \vspace{0.1in}
  \begin{tabular}{ccc}
    percolation type & $d_w$ & $d_s$ \\
    \tableline
    ordinary & $2.841 \pm 0.015$ & $1.329 \pm 0.010$ \\
    FC       & $2.843 \pm 0.015$ & $1.326 \pm 0.017$ \\
    IFC      & $2.857 \pm 0.015$ & $1.300 \pm 0.012$ \\
    \tableline
    ordinary &                   &                   \\
    (prior results) & $2.87 \pm 0.02$ 
    \tablenote{see Majid, et. al. \cite{majid84}.} & 
    $1.30 \pm 0.02$ \tablenote{see, e.g., Nakanishi \cite{nakanishi94}} \\
  \end{tabular}
  \vspace{0.1in}
  \end{table}

  Listed in the second column of Table~\ref{tab:dynamic} are the $d_w$ values 
  we found following Eq.~(\ref{walk_dim}) for ordinary, FC and IFC 
  percolation.  Notice that there is an excellent agreement among the values 
  for all three cases and that they agree with the previously known value for 
  ordinary percolation.

  In the third column of Table~\ref{tab:dynamic} are listed the $d_s$ values
  calculated using the Alexander-Orbach scaling law.  Although all three
  estimates from this work agree with the previously known value for
  ordinary percolation within the aggregate error estimates of our work
  and those of the previous work, there is a slight internal discrepancy
  between our own estimate of $d_s$ for ordinary percolation and that
  for IFC percolation. Since $d_s$ is derived from $d_f$ and $d_w$, the
  slight discrepancy may be attributed to the $d_f$ value we found for
  IFC percolation by direct simulation (refer to the third column of
  Table~\ref{tab:static}).  We suspect an unaccounted systematic error 
  occuring when we perform each iteration process. In particular, as we
  will discuss in more details later, the length scales associated with
  fully-coordinated subregions appear to increase at each iteration. 
  This suggests that we may have to increase the overall lattice length
  scale correspondingly in order to maintain a good estimate of the
  fractal dimension by direct simulation.  Since this was not done in this
  work, a systematic error may have crept in this way.

\section{GEOMETRY OF CLUSTERS}
\label{geometry}

\begin{figure}
  \epsfxsize=1.0\hsize \mbox{\hspace*{-0.065\hsize} \epsffile{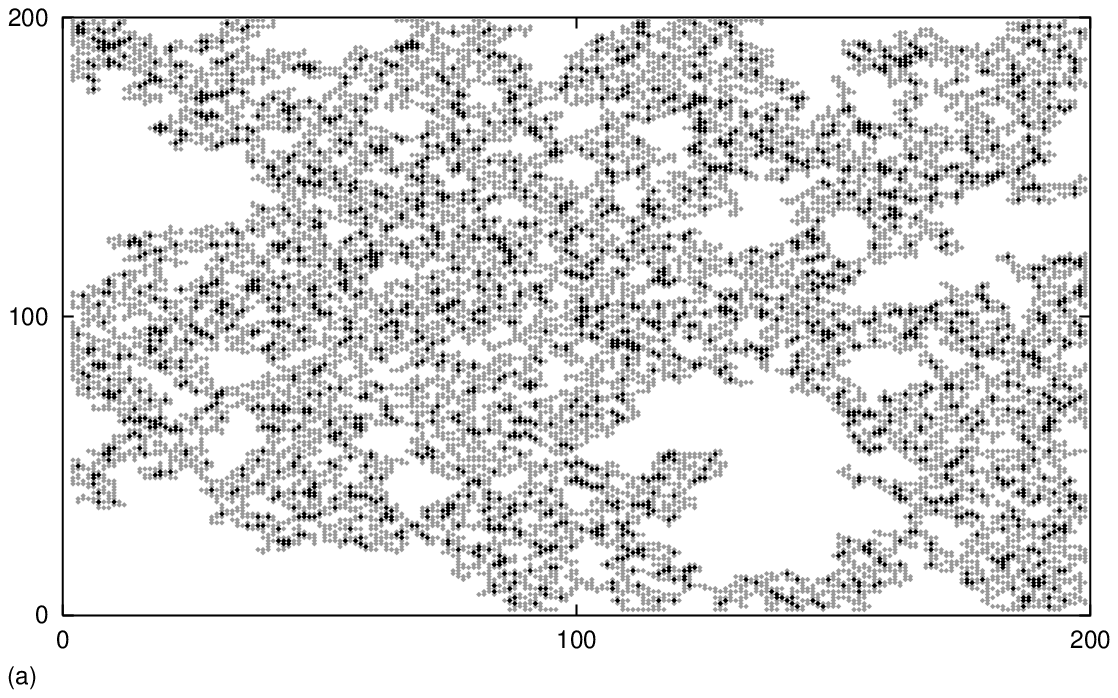}}
  \vspace{0.2cm}

  \epsfxsize=1.0\hsize \mbox{\hspace*{-0.065\hsize} \epsffile{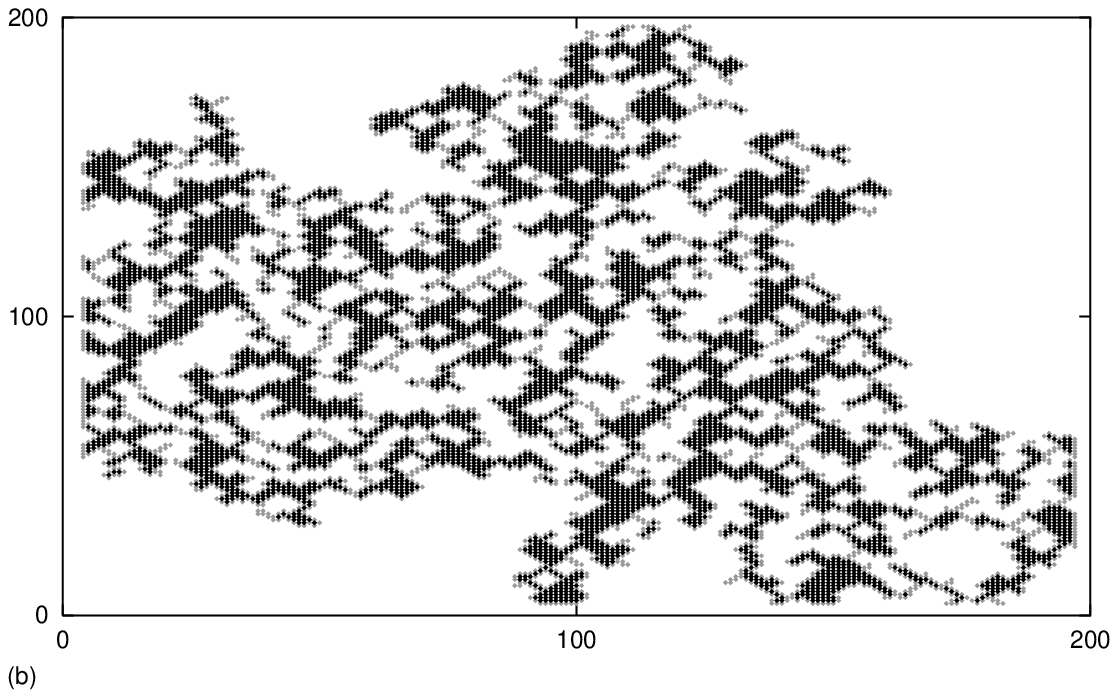}}
  \vspace{0.3cm}
  \caption{Sample largest clusters resulting from simulating ordinary and
           IFC percolation at their respective percolation thresholds in a
           $200$x$200$ square lattice with free boundaries are shown.  The 
           darker dotted sites are FC (fully coordinated) sites while the 
           lighter gray dotted sites are PC (partially coordinated) sites.  
           (a) For ordinary percolation, PC sites dominate.  The occasional 
           FC sites are clumped together and are somewhat distributed evenly 
           in the cluster.  (b) For IFC percolation, the number of FC sites
           increases.}
  \label{clusters}
  \end{figure}

  Previously \cite{fully}, it was found that the number of {\it interior} sites
  in the largest cluster in FC percolation is much greater than that in
  ordinary percolation at their respective thresholds. In this work we find 
  that this trend continues further for IFC percolation.  As an illustration, 
  consider Figs.~\ref{clusters}(a) and \ref{clusters}(b).  Shown in 
  Fig.~\ref{clusters}(a) is the largest cluster from a simulation of 
  ordinary percolation on a $200$x$200$ square lattice at its percolation 
  threshold.  The darker dotted sites are {\it fully-coordinated} (FC) sites 
  while the lighter gray dotted sites are {\it partially-coordinated} (PC) 
  sites. FC sites are those all of whose neighbors are also occupied (thus 
  these are similar to, though not identical with, the {\it interior} sites 
  considered in \cite{fully}), while PC sites have at least one nearest 
  neighbor not occupied.  The same shading tone is used in 
  Fig.~\ref{clusters}(b) except that what is shown is the largest cluster 
  from a simulation of IFC percolation at its own percolation threshold.  
  Notice the difference in the number of FC sites in these figures.  Partially
  coordinated sites dominate in ordinary percolation while FC sites dominate 
  in IFC percolation.

\begin{figure}
  \epsfxsize=0.90\hsize \mbox{\epsffile{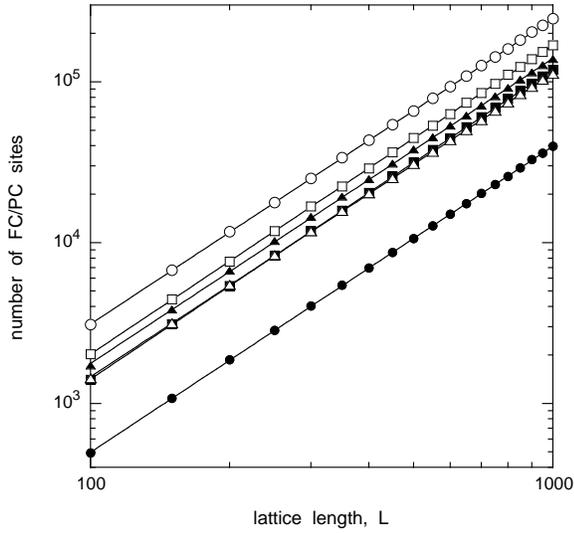}}
  \vspace{0.3cm}
  \caption{A log-log plot of the number of FC and PC sites in the largest
           cluster at the percolation threshold against the lattice length 
           $L$ is shown.  The circles are for PC sites while the solid circles 
           are for FC sites, both in ordinary percolation.  The data points
           for FC percolation are the squares (which are PC sites) and the
           solid squares (which are FC sites).  For IFC percolation, the 
           triangles are for PC sites while the solid triangles are for FC 
           sites.  Each data point is an average over $2000$ random 
           realizations.}
  \label{fcpc}
  \end{figure}

  The average number of FC and PC sites in ordinary, FC, and IFC percolation
  are plotted in Fig.~\ref{fcpc}.  Linear least squares fits indicate that in 
  ordinary percolation the PC sites are more than $6$ times as many as the 
  FC sites, while the ratio in FC percolation is about $1.5$, and in IFC 
  percolation it is only about $0.86$.  A trend that can easily be seen here 
  is that the number of FC sites increases as we progress from ordinary to 
  FC and then to IFC percolation.  Although only a local full-coordination 
  correlation is imposed, it clearly results in a change on the global scale. 
  This may be reflected in some qualitative differences in a physical property
  if it is sensitive to the size of connected clusters of FC sites. For 
  example, if a percolation model is used to describe a disordered substrate 
  on which diffusive physics takes place (such as in catalysis), what is 
  expected within a given time scale (or frequency range) would be directly 
  affected by the size of the high-connectivity or FC regions of the substrate
  where fast diffusion occurs compared to less well connected areas. In such 
  cases, it is expected that there would be large differences depending on
  whether the appropriate percolation model to use is ordinary, FC, or IFC 
  percolation. The latter cases could be a better representation if the 
  culling of partially coordinated sites naturally occurs in the particular 
  problem, say, due to desorption.

  \begin{figure}
  \epsfxsize=0.90\hsize \mbox{\epsffile{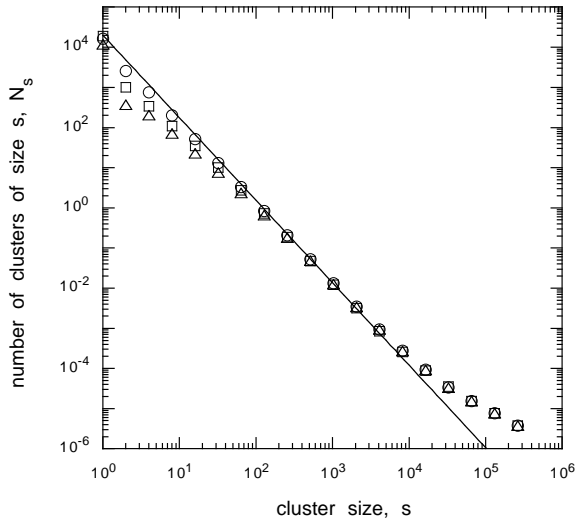}}
  \vspace{0.3cm}
  \caption{A distribution of cluster sizes is shown in a log-log plot.  The
           circles, squares and triangles are for ordinary, FC, and IFC
           percolation, respectively.  Each data point is an average over
           $2000$ random realizations.  The line drawn is not the best fitting
           line but instead one that follows $N_{s} \sim s^{-\tau}$.}
  \label{tau}
  \end{figure}

  \begin{figure}
  \epsfxsize=0.75\hsize \mbox{\hspace*{0.08\hsize} \epsffile{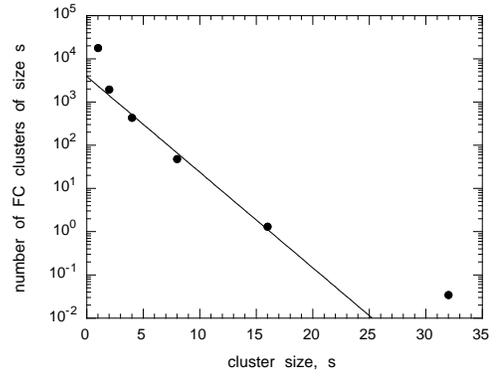}}

  \epsfxsize=0.75\hsize \mbox{\hspace*{0.08\hsize} \epsffile{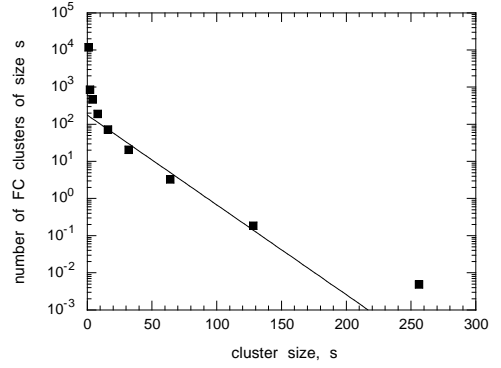}}

  \epsfxsize=0.75\hsize \mbox{\hspace*{0.08\hsize} \epsffile{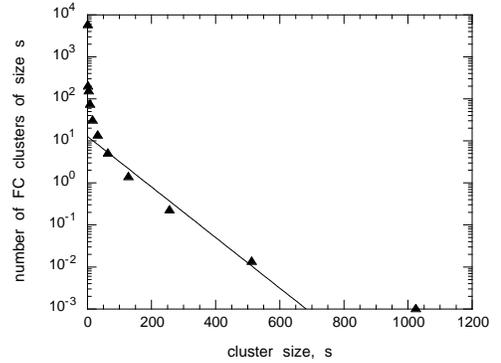}}
  \vspace{0.3cm}
  \caption{Shown are semi-log plots for the number of clusters formed by FC
           sites against the corresponding cluster sizes.  Each line drawn
           in each plot does not correspond to the best fit line for all
           available points but only for the four middle points. This would
           give a rough estimate of the average FC cluster sizes.  (a) In
           ordinary percolation, on average there are $2$ FC sites in each
           FC cluster.  (b) In FC percolation, on average FC clusters have
           $18$ FC sites.  (c) In IFC percolation, there are around $72$
           FC sites in an FC cluster.}
  \label{fc_clusters}
  \end{figure}

  The distribution of cluster sizes (of all sites, i.e., both FC and PC
  sites) is shown in Fig.~\ref{tau}.  Because of the finite size effects, 
  only the intermediate size ranges (about $10^2$ to $10^4$ sites) are 
  consistent with the cluster size scaling $N_{s} \sim s^{-\tau}$ with the 
  ordinary percolation value of $\tau = 187/91 \approx 2.055$, where $s$
  is the cluster size and $N_s$ is the number of clusters having size $s$.
  Very small $s$ values are clearly out of the asymptotic scaling regime 
  while the region of $s \sim {\cal{O}}(L^2)$ is also limited by the finite 
  size of the grid.  Nonetheless, it is important to note that there is no
  significant difference between ordinary, FC, and IFC percolation for all 
  $s$ above about $10^2$.  This suggests strongly that the asymptotic cluster 
  size scaling (which is known to occur for ordinary percolation) also occurs
  for both FC and IFC percolation.

  In making the plot for Fig.~\ref{tau}, we didn't distinguish between PC and
  FC sites in the clusters.  If we instead just consider FC sites and make a 
  semi-log plot of the number of clusters formed by these sites against the
  corresponding cluster sizes, we should be able to roughly estimate the 
  average cluster sizes formed by FC sites alone.  Shown in 
  Fig.~\ref{fc_clusters}(a) is $N_{s}$ against $s$ of clusters formed by FC 
  sites in ordinary percolation.  The line shown is not a best fitting line 
  for all of the available points but instead one that best fits only the 
  four middle points and is shown merely to obtain a crude idea of the typical
  size scale of FC clusters.  For ordinary percolation, we found the average 
  FC cluster size to be around $s_0 \approx 2$ FC sites.  Shown in 
  Figs.~\ref{fc_clusters}(b) and \ref{fc_clusters}(c) are data for FC
  and IFC percolation, respectively.  We found that around $s_1 \approx 18$
  FC sites comprise the average cluster in FC percolation while around
  $s_2 \approx 72$ FC sites make up the average FC cluster in IFC percolation.
  This means that in order to map the FC and IFC percolation problems to the
  same length scale as ordinary percolation, a length rescaling by
  $\sqrt{s_1/s_0} \approx 3$ and $\sqrt{s_2/s_0} \approx 6$ must be performed,
  respectively. Thus, although both static and dynamic critical exponents
  seem to remain the same as the full-coordination rules are iteratively
  applied, the local connectivity greatly changes or, equivalently, the
  length scale which would preserve the local connectivity changes.

\section{COMPARISON WITH PREVIOUS WORK ON FC PERCOLATION}
\label{comparison}

  In this work, we follow the prescription described in the Introduction
  closely in order to generate the ensemble of FC and then IFC clusters.
  The only limitation in the Monte Carlo portion of our work are that
  the grid we use to generate the clusters must be finite and that
  a particular boundary condition must be selected.  Our largest grid is
  $2048$x$2048$ and clearly, this is not very large by today's standards.
  Since we use finite size scaling in the determination of both the
  static and dynamic critical exponents, the finiteness of the grid (and
  cluster) is in fact necessary; yet, there is always a possibility that
  an asymptotic regime of large sizes may not have been reached.  Also,
  we use the free boundary condition with no direct connectivity across the
  opposite edges of the grid, which may accentuate the finite size limitations.
  This type of ensemble is the most straightforward extension of the standard
  Monte Carlo simulation of ordinary percolation to the FC and IFC
  percolation models, and may be described as {\em static} since there is no 
  dependence on the order in which sites are generated or the connectivities 
  checked.  Indeed, the static exponents $\gamma$, $\gamma/\nu$, and the like,
  agree closely with the corresponding ordinary static percolation values, 
  and, if the length scale is identified with the grid edge length, then the 
  cluster's fractal dimensions should also agree closely with that of 
  ordinary percolation.

  \begin{figure}
  \epsfxsize=0.75\hsize \mbox{\hspace*{0.08\hsize} \epsffile{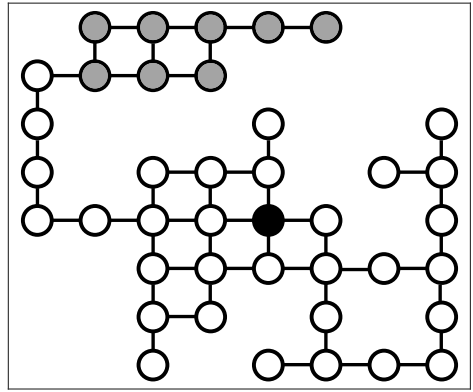}}
  \vspace{0.3cm}
  \caption{Figure to illustrate cluster geometry differences between growing
           a cluster from a seed and generating it in a square lattice (static
           ensemble).  The solid circle is the seed and the cluster grown
           from it includes all the circles but excluding the gray-filled
           ones.  The growth process stops when a specified cluster size is 
           reached.  In this figure, the gray-filled circles are excluded
           because the specified cluster size has already been reached; 
           however, when generating the cluster in a square lattice of the 
           same length scale, the gray-filled circles will be included in 
           the cluster.}
  \label{generate}
  \end{figure}

  In our previous work \cite{fully}, we studied the static and dynamic
  critical behavior of FC percolation using the ensemble of clusters
  of specified number of sites, say, $S'$. Those clusters were grown from
  a seed site by applying a {\em breadth-first} algorithm twice; i.e.,
  first generating an ordinary percolation cluster in nearest neighbor
  shells to a sufficiently large size (say, $5S'$), and then identifying
  and joining only the fully coordinated sites in a second {\em searching}
  process from the same seed and then stopping the process when the size of
  the resulting connected cluster reaches the predetermined value $S'$. 
  Now, if we make the intermediate cluster of the first search process
  very large and if we do {\em not} terminate the second search process
  arbitrarily when $S'$ is reached, but rather continue the search until
  the natural boundary of the FC cluster is found, then the resulting cluster
  ensemble should be {\em exactly} the same as that which would result in
  the {\em static} ensemble of FC percolation clusters generated on an
  {\em infinite} grid.  However, since the first stage is stopped at 
  at most $5S'$ and since the second stage is forcibly stopped exactly at
  $S'$, which is a predefined value, we introduced a kinetic nature to the 
  cluster ensemble.  This may be somewhat similar to the differences in the 
  degree of intrinsic anisotropy in static and {\em kinetic} ensemble of 
  percolation clusters studied previously \cite{family,jacobs} (called 
  {\em equlibrium} versus {\em growing} clusters there).

  An illustration of the cluster geometry differences arising between a static 
  cluster and a growing one is shown in Fig.~\ref{generate}.  In the static 
  ensemble, every site of the spanning, connected component of the ordinary, 
  FC, or IFC percolation models' realizations within a given length scale $L$ 
  (defined to be the grid edge length) is a part of the cluster.  In the 
  growing cluster, on the other hand, only those sites which are within a 
  certain path length along the nearest neighbor connections from the seed 
  site are counted, independent of how close they are by straight line 
  distance to the seed.  Also, since cluster size $S'$ is the control 
  parameter for this ensemble, the length scale $L$ is not precisely fixed 
  but rather obtained as an average for a given value of $S'$.  Exactly which 
  sites are counted in a cluster of length scale $L$ and which ones excluded 
  do clearly depend on the particular search order that is employed.  Thus, 
  typically, for the same $L$, the number of sites of the percolating cluster 
  is smaller in the growing cluster ensemble than in the static one.  This 
  tends to reduce both the fractal dimension $d_f$ and the walk dimension 
  $d_w$.  If the effect is more pronounced for $d_w$, then it may explain why 
  an apparent measurement of the spectral dimension $d_s = 2 d_f/d_w$ was 
  larger for FC in the growing cluster ensemble.  Since the search algorithm 
  is repeated to obtain an IFC cluster, the effect will be even larger in IFC 
  than in FC percolation, as well as it is larger in FC than in ordinary 
  percolation.

\section{SUMMARY}
\label{summary}

  In summary, we have introduced and studied an extension to ordinary site
  percolation on a square lattice which we call {\it iterated fully
  coordinated percolation}.  The static and dynamic critical behaviors of
  this model, as well as that of the ordinary and {\it fully coordinated}
  percolation models, were determined using methods such as Monte Carlo 
  simulations, finite-size scaling, and normal mode analyses.  All clusters
  used in our simulations were generated statically in grids of predetermined
  sizes on a square lattice, as opposed to growing them dynamically from a
  seed according to a dynamic rule.  We found that the sizes of subregions of 
  full connectivities in an incipient infinite cluster greatly increases as
  we simulate from ordinary to fully coordinated and then to iterated fully 
  coordinated percolation.  The length scales of such subregions increase by
  a factor of two to three with each iteration. Although this implies a large
  increase in the local connectivities, our results also indicate that the
  ordinary, fully coordinated, and iterated fully coordinated percolation
  models have identical static and dynamic critical exponents within our
  statistical errors and thus we believe that they belong to the same
  universality class, both statically and dynamically.


\end{multicols}
\end{document}